# An analysis of overall network architecture reveals an infinite-period bifurcation underlying oscillation arrest in the segmentation clock

Zavala E[a], Santillán M[b, c]

[a] *Centro de Investigación y de Estudios Avanzados del IPN, Depto. de Biomedicina Molecular. Av. Instituto Politécnico Nacional 2508, Col. San Pedro Zacatenco, CP 07360 México DF, México*
[b] *Centro de Investigación y de Estudios Avanzados del IPN, Unidad Monterrey. Vía del Conocimiento 201, Parque PIIT, CP 66600 Apodaca NL, México*
[c] *Centre for Applied Mathematics in Bioscience and Medicine. 3655 Promenade Sir William Osler, McIntyre Medical Building, Room 1123A, Montreal, QC H3G 1Y6, Canada*

## Abstract

Unveiling the mechanisms through which the somitogenesis regulatory network exerts spatiotemporal control of the somitic patterning has required a combination of experimental and mathematical modeling strategies. Significant progress has been made for the zebrafish clockwork. However, due to its complexity, the clockwork of the amniote segmentation regulatory network has not been fully elucidated. Here, we address the question of how oscillations are arrested in the amniote segmentation clock. We do this by constructing a minimal model of the regulatory network, which privileges architectural information over molecular details. With a suitable choice of parameters, our model is able to reproduce the oscillatory behavior of the Wnt, Notch and FGF signaling pathways in presomitic mesoderm (PSM) cells. By introducing positional information via a single Wnt3a gradient, we show that oscillations are arrested following an infinite-period bifurcation. Notably: the oscillations increase their amplitude as cells approach the anterior PSM and remain in an upregulated state when arrested; the transition from the oscillatory regime to the upregulated state exhibits hysteresis; and an opposing distribution of the Fgf8 and RA gradients in the PSM arises naturally in our simulations. We hypothesize that the interaction between a limit cycle (originated by the Notch delayed-negative feedback loop) and a bistable switch (originated by the Wnt-Notch positive cross-regulation) is responsible for the observed segmentation patterning. Our results agree with previously unexplained experimental observations and suggest a simple plausible mechanism for spatiotemporal control of somitogenesis in amniotes.

keywords: somitogenesis, vertebrate embryo, delay differential equations, bistability, SNIC/SNIPER bifurcation

# Introduction

Vertebrate segmentation is an embryonic developmental process that starts just after gastrulation and occurs along the presomitic mesoderm (PSM) of the embryo. It consists of the rhythmic and sequential formation of cell blocks, the somites, which become the precursors of vertebrae, ribs, intercostal muscles and other associated tissues of the thorax [6, 20]. Because of this, the process is also known as somitogenesis. While immature mesenchymal cells are constantly added to the posterior PSM in a highly proliferative zone called the tail bud, somites arise in pairs from presomitic cells that cluster together in the anterior PSM. In this way, the somite boundaries appear progressively in an antero-posterior (AP) direction, starting near the otic vesicle and ending at the tail bud of the embryo. The somite number and periodicity can vary among species, but are highly conserved within a given species.

Since the first observations of somitogenesis by Malpighi, efforts have been made to understand its regularity and high similitude in all vertebrate species. In 1976, Cooke and Zeeman [8] introduced the *Clock and Wavefront* model in an attempt to qualitatively explain how the interaction of a clock mechanism with an AP traveling maturation wavefront could be responsible for the segmental patterning. However, it was not until 1997 [24] that the first molecular components of a genetic oscillator were identified in the PSM of chick. Since then, many genes with oscillatory expression have been associated to somitogenesis in different vertebrate species. For instance, approximately 30 genes with periodic expression have been identified in the mouse PSM [9]. These genes synchronize its expression within nearest neighbor cells and generate a wave-like expression pattern which starts at the tail bud of the embryo, travels anteriorly, slows down while approaching the anterior PSM and stops abruptly at the site where the future somite pair will form. The period of these *clock-waves* matches the formation rhythm of somites. The gene expression pattern resulting after oscillation arrest precedes the maturation stage, which involves the activation of segmentation genes, like *Mesp2*, that control somitic cellular identity and polarity [27, 10].

The question of what the clock mechanism that drives oscillations in PSM cells is remains elusive. Significant progress has been made using the zebrafish as experimental model [23, 25, 4]. However, the molecular components controlling somitogenesis in amniotes such as chick and mice are much more complex. In mice, the cyclic genes are known to be downstream of either the Wnt, Notch, or FGF signaling pathways. Predominance and redundancy of negative feedback at multiple levels are perhaps the most remarkable features of the internal wiring found within each pathway [10]. Previous mathematical models and experiments in zebrafish strongly suggest that delayed negative feedbacks are responsible for self-sustained oscillations [23]. However, in amniotes, the abundance of cross-regulation among the three pathways at many levels makes it difficult to assert that the expression dynamics of each gene is exclusively determined by its upstream pathway [10, 15].

The Wnt, Notch, and FGF signaling pathways are linked with spatiotemporal gradients known to work as positional cues to PSM cells [12, 7]. There are four biomolecules that exhibit graded concentration profiles across the PSM: Retinoic Acid (RA) [11], *Wnt3a* [2], nuclear β-catenin [3], and Fgf8 [13]. RA distribution follows a classical morphogenetic gradient, established by a source-sink mechanism. It has its maximum concentration in the somites an anterior PSM, and diminishes posteriorly. Conversely, *Wnt3a*, nuclear β-catenin, and Fgf8 have gradient distributions opposing that of RA. *Wnt3a* and nuclear β-catenin are, respectively, the ligand and the canonical intracellular mediator of Wnt signaling. Fgf8 is the ligand of receptor Fgfr1 (the only FGF receptor expressed in the PSM [29]) and the activator of FGF signaling. In these cases, the gradient profile is stablished because all three substances are highly expressed in the tail bud and posterior PSM, and their anterior concentration diminishes due to protein and mRNA turnover during embryo extension [1]. It is important to point out that the experimental evidence supports that all substances with a gradient distribution are downstream of Wnt signaling, although the participation of other pathways in its regulation has not been ruled out [1]. Finally, Fgf8 and RA are involved in a double-negative feedback circuit of antagonistic regulation. Mathematical modeling of this Fgf8/RA circuit has suggested that a traveling bistability domain may be responsible for the abrupt transition of PSM to somitic cells [18].

The involvement of Wnt and FGF signaling in both establishing the gradients and maintaining oscillations is not yet well understood. However, some conjectures can be made from mathematical modeling studies addressing these questions. Santillán et al. [28] introduced a minimal regulatory network comprising Notch an Wnt activities, and were able to provide a potential explanation for the periodic aggregation of PSM cells into blocks, which is a necessary step for the maturation of somites. In that model, Notch self-regulation was represented as a negative-feedback circuit. This circuit was further coupled to Wnt through positive bidirectional crosstalk. This simple network predicted self-sustained oscillations of PSM cells, and variations on a bifurcation parameter, representing the positional information of a hypothetical gradient, arrested oscillations sequentially following a somitic pattern. On the down side, this model failed to explain the increment of the PSM-cells cycling-period as they approach the anterior PSM.

In the present work we address the question of how oscillations are arrested in the amniote segmentation clock, from a mathematical modeling perspective. In consequence, we introduce a minimal model that accounts for the crosstalk between the Wnt, Notch and FGF pathways, together with the negative cross-regulation of Fgf8 and RA. To our knowledge, this is the first model that takes into consideration the mutual interaction between all three regulatory pathways [18, 19, 22, 26, 28]. Furthermore, although semi-quantitative, our model is biologically inspired on the structural information available on the mouse somitogenesis regulatory-network. Finally, despite its simplicity, this model is capable of reproducing most of the dynamical features of the oscillation-arrest phenomenon: PSM cells show sustained oscillations, the clock-wave slows down and its oscillation amplitude increases as it travels anteriorly, and oscillations are irreversibly arrested at the anterior-most PSM. In this respect it is important to emphasize that the cause underlying the slowing down of oscillations previous to its definitive arrest has remained elusive since first observed in 1997 [24].

## Methods

*Model Development*

The mouse somitogenesis regulatory network can be visualized as an assembly of negative and positive circuits that build up the global network [10, 15]. Delayed negative and positive feedback circuits are broadly studied motifs known for their capability of, respectively, behaving as an oscillator and a bistable switch. Although negative feedback is suspected to be behind the oscillatory dynamics of various signaling pathways, in the context of somite formation it has only been extensively studied in the Notch pathway [15]. Special attention to Notch is perhaps due to historical reasons: the firstly identified clock genes are downstream this pathway. However, its importance also relies on its conserved role in vertebrate segmentation among different species. We thus decided to consider the Notch pathway as the only source of oscillations and modeled it as a delayed negative feedback circuit. After carrying out an extensive search in the literature (see Supporting Table S1 and Supporting Figure S1), we included positive circuits in the crosstalk between Wnt and Notch, and between Wnt and FGF; as well as a negative circuit in the crosstalk between the Notch and Fgf pathways. Finally, following Goldbeter et al., we added to our model the positive circuit created by the Fgf8 and RA mutually negative cross-regulation [18]. A schematic representation of the interactions accounted for in this model is given in Fig. 1.

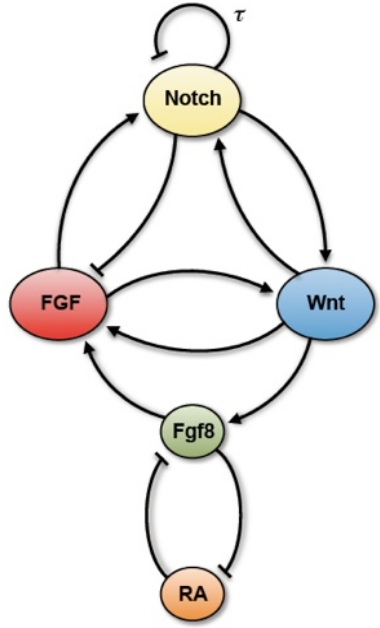

**Fig. 1** Somitogenesis regulatory network architecture considered in the model. Positive and negative interactions are represented, respectively, by arrows and T lines. The $\tau$ corresponds to the delay associated to Notch self-regulation.

The corresponding model equations are:

$$\begin{aligned}
\dot{w} &= a_w + g_{wn}(n) + g_{wf}(f) - \gamma_w w, \\
\dot{n} &= a_n + h_{nn}(n_\tau) + g_{nw}(w) + g_{nf}(f) - \gamma_n n, \\
\dot{f} &= a_f + g_{fw}(w) + h_{fn}(n) + \beta_{fx} x - \gamma_f f, \quad (1)\\
\dot{x} &= a_x + g_{xw}(w) + h_{xy}(y) - \gamma_x x, \\
\dot{y} &= a_y + h_{yx}(x) - \gamma_y y.
\end{aligned}$$

Rather than accounting for specific gene expression levels, variables $w$, $n$, and $f$ respectively represent the Wnt, Notch, and FGF global pathway activities. Conversely, $x$ and $y$ are proportional to the Fgf8 and RA concentrations. For simplicity, all of these variables are assumed to be non-dimensional. Parameters $a_i$ and $\gamma_i$ ($i = w, n, f, x,$ and $y$) respectively stand for the basal input rates and the turnover rates of the corresponding variables, while $\tau$ is the time delay associated to the Notch negative feedback circuit. This delay stands for the sum of the duration of processes like transcription, mRNA processing, translation, post-translational modifications and intracellular transport of signaling molecules. Finally, functions $0 \leq g_{ij}(j) < \beta_{ij}$ and $0 < h_{ij}(j) \leq \beta_{ij}$ represent the regulation exerted by variable $j$ upon variable $i$. Functions $g_{ij}$ correspond to positive regulations, while functions $h_{ij}$ correspond to negative regulations (see Fig. 1). We have assumed that $g_{ij}$ and $h_{ij}$ are, respectively, increasing and decreasing Hill type functions of the form

$$g_{ij}(j) = \beta_{ij} \frac{j^4}{K_{ij}^4 + j^4},$$
$$h_{ij}(j) = \beta_{ij} \frac{K_{ij}^4}{K_{ij}^4 + j^4}, \quad (2)$$

where the $\beta_{ij}$'s represent the corresponding regulation weights, while the $K_{ij}$'s denote the half saturation constants. For simplicity, all Hill coefficients were set to 4 to assure enough non-linearity while still falling within a biological plausible level of cooperativity. The only exception for non-linear regulation is the regulatory influence of Fgf8 on the FGF dynamics. This was modeled as a linear function because it is the simplest way to describe the binding of the Fgf8 ligand to its Fgfr1 receptor.

Notice that the influence of each state variable on the others was modeled by considering additive regulatory functions instead of multiplicative ones. This assumption is supported by the fact that, in eukaryotes, although the so-called essential transcription factors are absolutely necessary for a polymerase to bind a given promoter and start transcription, there are also non-essential transcription factors, known as activators and coactivators, which can combine to provide a diversity of responses to various signals. This form of regulation is incompatible with multiplicative regulatory inputs, because all of them need to be present so transcription can be started, but is in complete agreement with additive regulatory inputs.

Since we aim at emphasizing the importance of the network architecture in determining the dynamical behavior of the segmentation clock, we consider that the minimal model introduced above is simple enough to be thoughtfully analyzed, while still capturing the essential biological characteristics of the whole regulatory network.

*Numerical Methods*

The numerical solutions and bifurcation analysis of our model were carried out using the software XPPAUT and the numerical methods available in it [14]. The simulations of sequential clock-waves along a linear array of PSM cells were carried out using MATLAB.

# Results

We started by studying the effects of separately varying the parameter values within the negative and positive circuits. We found for the Notch delayed negative self-regulation that a strong feedback ($K_{nn} \approx 0.3$) and a long delay ($\tau \approx$ 45 min) are necessary to observe self-sustained oscillations. Conversely, from the three positive circuits that can be identified in the network (see Fig. 1), we chose to explore the bistable potential of the circuit formed by the bidirectional crosstalk between Wnt and Notch (this follows a previous work predicting bistable properties for this circuit [28]). Thus, we assigned this circuit a high influence with respect to other positive circuits in the network. We did this by setting a strong weight to its cross-regulatory functions ($\beta_{wn} = \beta_{nw} = 0.2$ min$^{-1}$) compared to others. We set the values of the time delay and the turnover rates in order to reproduce the observed ~120 min period of clock-wave oscillations in mouse somitogenesis. The remaining parameters were adjusted so the oscillations in the three pathways rapidly converge to a stable orbit and their amplitudes are comparable. The resulting nominal values are listed in Table 1.

**Table 1**. Nominal parameter values.

| Parameter | Nominal value |
|---|---|
| $a_w, a_n, a_f, a_x, a_y$ | $0.02\ \text{min}^{-1}$ |
| $\beta_{nn}, \beta_{xw}, \beta_{xy}, \beta_{yx}$ | $0.1\ \text{min}^{-1}$ |
| $\beta_{wn}, \beta_{nw}$ | $0.2\ \text{min}^{-1}$ |
| $\beta_{wf}, \beta_{fw}$ | $0.06\ \text{min}^{-1}$ |
| $\beta_{nf}, \beta_{fn}$ | $0.04\ \text{min}^{-1}$ |
| $\beta_{fx}$ | $0.02\ \text{min}^{-1}$ |
| $K_{nn}$ | $0.3$ |
| $K_{wn}, K_{nw}, K_{wf}, K_{nf}, K_{xy}, K_{yx}$ | $1$ |
| $K_{fw}$ | $0.4$ |
| $K_{fn}$ | $1.5$ |
| $K_{xw}$ | $0.1$ |
| $\gamma_w, \gamma_y$ | $0.14\ \text{min}^{-1}$ |
| $\gamma_n$ | $0.15\ \text{min}^{-1}$ |
| $\gamma_f, \gamma_x$ | $0.16\ \text{min}^{-1}$ |
| $\tau$ | $45\ \text{min}$ |

The self-sustained oscillations predicted by our model when using the parameter values in Table 1 correspond to an hypothetical presomitic cell that has not yet abandoned the tail bud. As the embryo keeps growing, these cells are left behind and undergo a relative displacement to more anterior regions of the PSM. In the process, they sense a decreasing concentration of *wnt3a*, which translates to Wnt3a, the ligand that activates the Wnt signaling pathway. This pathway in turns modulates other gradients like nuclear β-catenin, Fgf8 and, indirectly, RA (see Supporting Fig. S1 and Supporting Table S1). Hence, we simulated the effect of a Wnt3a gradient on PSM cells by introducing a new parameter, *k*, which represents the effects of a decreasing Wnt signal in two ways: firstly, by modulating the half-life of the Notch1-receptor intracellular domain NICD (NICD is released after binding of the Notch1 ligand Dll1, together with Rbpj-κ it activates transcription of genes downstream of Notch, and it is known to be destabilized by Wnt signaling [2]); and secondly, by modulating transcription of *Fgf8*, which is know to be downstream Wnt signaling and expressed only in the posterior PSM. In this way, the Wnt3a gradient affects simultaneously the oscillatory and bistable circuits, as well as the antagonistic Fgf8/RA gradients. We represent this effects mathematically by replacing the corresponding parameters $\gamma_n$ and $\beta_{xw}$ in Eq. 1 according to the following rules:

$$\gamma_n = \left[(0.02)k + 0.13\right]\text{min}^{-1},$$
$$\beta_{xw} = \left[(0.1)k\right]\text{min}^{-1}, \qquad (3)$$

with *k* varying from one down to zero. Although the specific mechanisms by which the clock and gradients interact are poorly understood, we found that the simple effects considered in Eq. 3 account for both the opposed distribution of Fgf8 and RA gradients, as well as for the increasing period of oscillations in PSM cells prior to their arrest. To corroborate this, we made a bifurcation analysis of the model using *k* as the bifurcation parameter. We started at *k* = 1, which corresponds to an immature cycling cell in the tail bud. At this point, the cell experiences the maximum Wnt3a activity. As the embryo develops and the tail bud recedes, the cell enters the PSM and lingers on it until it is eventually reached by the maturation wavefront and becomes part of a somite. In the process, the cell senses lower values of Wnt3a as its relative position moves from the posterior to the anterior PSM. We modeled this by lowering the value of parameter *k*. The resulting bifurcation diagrams for the three pathways as well as the Fgf8 and RA levels are shown in Fig. 2.

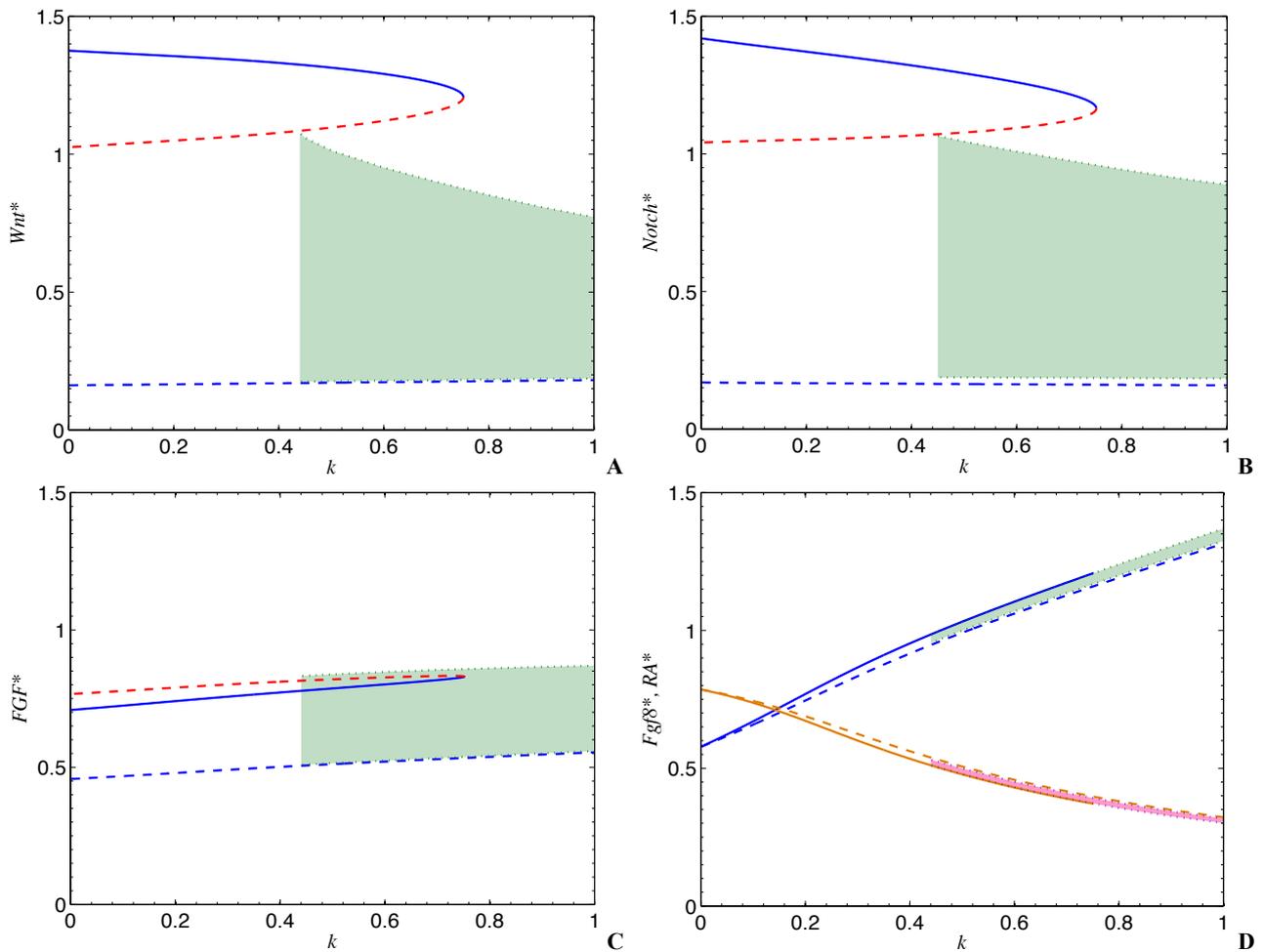

**Fig. 2** Bifurcation diagrams of **A**. the Wnt, **B**. the Notch, **C**. the FGF pathways and **D**. the Fgf8/RA opposing gradients. Oscillatory domains are shown in green. Stable steady-states are illustrated by continuous lines. Unstable steady-states are illustrated as dashed lines. Blue dashed lines denote the steady-state that becomes unstable because of the delay. In **D**, the oscillatory domain for RA and its steady-states are shown in pink and orange, respectively.

The most notorious characteristic of the bifurcation diagrams in Fig. 2 is the arrest of oscillations (shaded regions) observed at $k \approx 0.44$. Below this threshold, the system reaches a steady state characterized by high levels of Wnt and Notch activities, and a moderate FGF activity level. This steady state persists until $k = 0$. Notably, if the value of $k$ is now increased up from zero, the oscillatory behavior does not appear until the bifurcation parameter is above $k \approx 0.76$. That is, we observe the hysteretic behavior characteristic of bistable switches, although in this case one of the stable 'branches' corresponds to an oscillatory state. Another important result is the increment of the Wnt and Notch oscillation amplitude as $k$ decreases. This result is in agreement with experimental evidence regarding the expression level of genes under the Notch pathway as the PSM cells approach the differentiation front [2, 21, 30].

We also found that the arrest of oscillations occurs through an infinite-period bifurcation. This means that the period of oscillations increases dramatically as parameter $k$ approaches its critical value: $k \approx 0.44$. At this point, the period goes to infinite or, conversely, the frequency falls to zero (see Fig. 3A). This behavior has been reported earlier in the clock-wave of zebrafish [17] and chick [24]. In mice and chick, it has been suggested that the segmentation clock periodicity is regulated by Wnt [16]. Interestingly, Gibb et al. [15] suggested that a possible control mechanism of periodicity could be found in the interplay between Wnt and Notch. Specifically, by regulating the stability of one or more of the key clock components, just as in Eq. 3.

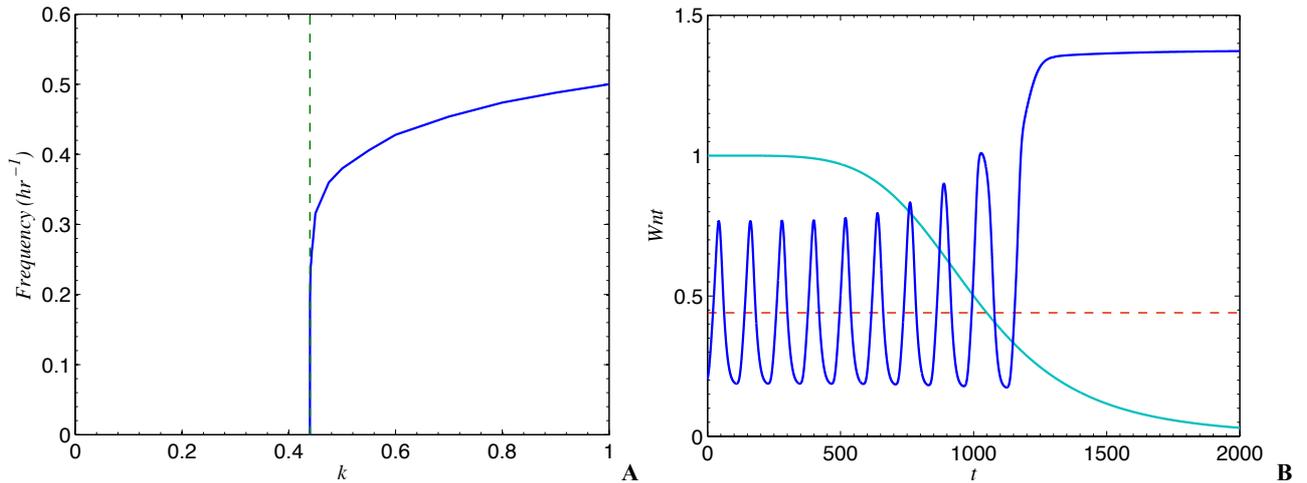

**Fig. 3** An infinite-period bifurcation underlies the arrest of oscillations. **A**. Frequency of oscillations drops to zero as parameter $k$ approaches its critical point at threshold value. **B**. Time evolution of *Wnt* oscillations (blue line) in a PSM cell modulated by a monotonic decreasing function $k(t)$ (sigmoid curve in green). The oscillations increase their amplitude and period as $k$ decreases. The arrest of oscillations is imminent once $k$ drops below its critical value (red dashed line).

Our results also suggest that the Fgf8 and RA levels are affected not only by their mutual antagonism, but also by the ceasing of *de novo* transcription of *Fgf8*. As mentioned before, *Fgf8* is a target under the Wnt pathway, which in turn is activated by Wnt3a and starts decreasing after PSM cells leave the tail bud. Our model reflects their opposing distribution in Fig. 2D. We can see there that Fgf8 decreases while RA increases as $k$ goes from 1 to 0. An oscillatory domain can be observed in the same interval as in Figs. 2A, 2B and 2C. This is an inevitable consequence of the coupling with the oscillating pathways. However, the oscillation amplitude is negligible, compared to that of the signaling pathways. We suggest that this low oscillation amplitude could be easily overshadowed by the biochemical noise present in the PSM environment.

We next evaluated how oscillations are affected by the bifurcation parameter in real time. To this end, we assumed that parameter $k$ decreases nonlinearly with time, affecting the time-course of oscillations in one PSM cell (see Fig. 3B). Our results show that as soon as the cell abandons the $k \approx 1$ zone at the tail bud and posterior PSM, its oscillations start increasing in amplitude and period. The stable steady-state of up-regulation is reached one cycle after the bifurcation parameter falls under $k \approx 0.44$. Once below this threshold, the arrest of oscillations is irreversible (at least while $k$ is kept below the upper threshold $k \approx 0.76$).

To better understand the meaning of all these single-cell results in the macroscopical context of the PSM, we implemented a simulation of the clock-wave. We followed the time evolution of 16 cells along the PSM, separated a fixed distance, in a linear array. Considering that the tail bud recedes at constant velocity and leaves cells behind steadily, the distance between two PSM cells in our array is proportional to the difference of elapsed times since their leaving the tail bud. We set this time difference to 30 min, so our array spans a PSM region four periods long, given an oscillation period of 120 min. To visualize the clock-wave along our cell set we assumed that the function describing how $k$ decays in time is delayed proportionally to the elapsed time since a given cell leaves the tail bud. That is, if $k(t)$ describes the time evolution of parameter $k$ for the anterior-most cell, $k(t - (i - 1)\Delta T)$ is the corresponding function for the $i$-th cell, with $\Delta T = 30$ min.

The resulting simulation of the clock-wave is shown in the Movie S1 of the Supporting Material. In Fig. 4 we present, from top to bottom, six sequential snapshots taken every 60 min of the state of all 16 cells. These snapshots mimic the classical *in situ* hybridization experiments aimed at detecting gene expression profiles in the PSM of vertebrate embryos [24]. There, we can identify the posterior-to-anterior propagation of the clock-wave (Fig. 4A), its narrowing as it approaches the anterior PSM (Fig. 4B), and its final arrest within a region that spans the size of a somite (which, in this case, is four cells width) (Fig. 4C). Subsequent clock-waves continue to arrive from the posterior PSM, reaching its final destination with a time difference of 120 min (Fig. 4C to 4F). This process can be more clearly visualized in the Movie S1 of the Supporting Material, where the simulation starts when all cells are located at the tail bud ($k = 1$) and their oscillations are synchronized. As time passes and cells sequentially leave the posterior PSM, their oscillation period increases until finally arrested to an up-regulated state at the anterior PSM. Importantly, oscillation arrest takes place in cohorts of four cells. The simulation ends when the oscillations in all cells are arrested and they have become committed to be part of a somite. Given that this commitment occurs sequentially in cohorts of cells that span the size of a somite, the maturation stage of somites, which follows the arrest of oscillations, will follow the same pattern as in our simulation. The maturation stage involves changes in cellular identity and polarity, as well as in the adhesive properties of cells that have become competent to segment [27]. These processes eventually make the intersomitic frontiers physically observable with a microscope (see Fig. 5). However, since these latter events are unlikely to be part of the clockwork controlling oscillation, slowdown, and arrest, they are not addressed here.

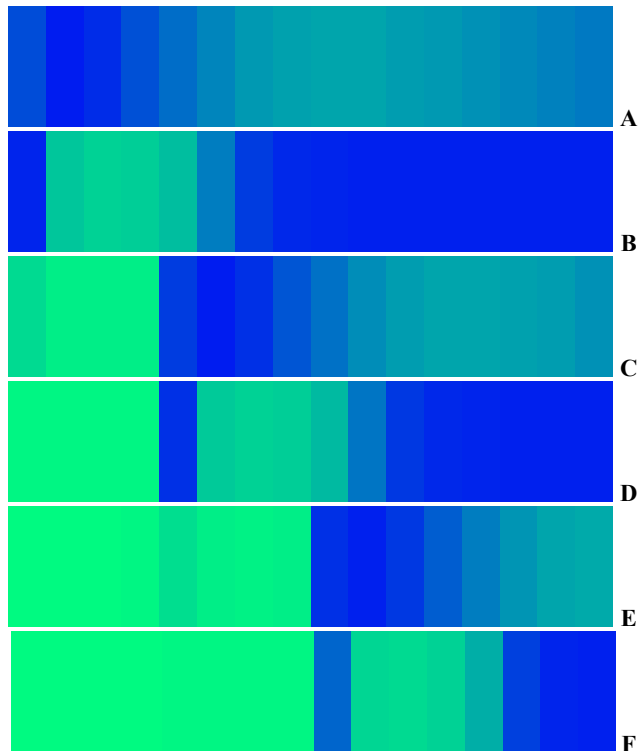

**Fig. 4** Snapshots, taken every 60 min, of a 16 cell linear array representing the PSM. The clock-waves start from tail bud and posterior PSM (right) and displaces anteriorly until arrested at future somite location (left) in cohorts of four cells. **A**. Propagation stage. **B**. Clock-wave slows down and narrows at arrival to anterior PSM. **C**. Arrested clock-wave spans somite size. Next clock-wave can be observed traveling from the right. **D**, **E** and **F**. The process is repeated with a periodicity of 120 min.

As previously discussed, we put a strong weight to the interactions between the Notch and the Wnt pathways. This opens the question of whether the same results can be obtained if the FGF, Fgf8, and RA nodes are eliminated from the network sketched in Fig. 1. We repeated the above-discussed bifurcation analysis with the modified model and show the results in Figs. S2 and S3 of the Supporting Material. We can observe there that the oscillatory regime disappears as the bifurcation parameter decreases below a given threshold and that this takes place via an infinite period bifurcation. Nevertheless, no hysteresis is observed whatsoever. The hysteretic behavior is important because it explains the experimentally observed irreversibility of PSM cells commitment. For instance, in a very nice experiment, committed cells were removed from chicken embryos a few minutes after oscillation arrest, and they were reinserted into the PSM, in a region where sustained oscillations of presomitic cells are observed. However, the reinserted cells remain in their non-oscillatory state—see Ref. [12] and references therein. In other words, hysteresis warranties that cells were oscillations have been arrested do not return to the oscillatory state due to fluctuations on the local Wnt3a concentration.

To test the robustness of the model we performed a sensitivity analysis of the periodic solution to variations in the parameter values (see Supporting Table S2). While developing the model we fixed the nominal parameter values in a range that preserves the self-sustained oscillations described before. This means that, although the results described here correspond to the nominal parameter values listed in Table 1, the model is robust enough to variations in these parameters so the oscillatory solution is still observed. Minimal variations were observed while deviating from the nominal parameter values, but they correspond to the shape of the periodic solution, its amplitude, and the position of the stable stationary-state (see Supporting Table S2). Moreover, the finding that oscillation arrest occurs via an infinite-period bifurcation and the reproduction of the segmentation pattern are preserved.

## Concluding remarks

Based on a minimalistic approach to the somitogenesis regulatory network, we have developed a simple mathematical model that accounts for the coupling of the Wnt, Notch and FGF signaling pathways, together with the Fgf8/RA mutually inhibitory circuit. Our model leaves aside almost all molecular details, but in exchange takes into account the mutual interactions between different regulatory pathways in a coarse-grained fashion. We considered Hill type additive regulatory functions between the model variables and assumed that the coupling between a delayed-negative and a positive feedback circuits is the most influential for reproducing the behavior of the clock-wave. To study the arrest of oscillations we introduced a bifurcation parameter corresponding to a Wnt graded signal. Although the lack of detail in our model prevented us from expecting quantitative predictions, it provides a simple and feasible explanation for the coupling between the gradient signal and the clock mechanism.

Our model results agree with numerous important experimental observations (see Fig. 5):

- It can reproduce oscillations in the three signaling pathways without assuming any external oscillatory input. That is, by taking the Notch internal negative feedback as the source of periodic behavior, the whole system oscillates spontaneously, as long as parameter $k$ takes values close to 1, which correspond to the cell being in the posterior PSM.

- Neighboring cells remain synchronized while they are in the tail bud and in the posterior PSM. After they leave this region, their oscillation period increases with time causing the clock-wave to narrow as it moves anteriorly.

- Once in the anterior PSM, the cells stop oscillating in equally sized cohorts. This is a strong requirement for generating equally sized somites (see Figs. 4 and 5). Remarkably, oscillation arrest occurs by means of an infinite-period bifurcation, which is consistent with the slowing down of the clock-wave prior to its final arrest in amniotes (a phenomenological description represented by the phase oscillator model of Campanelli has already provided a proof of concept [5]).

- A high expression state is observed after oscillation arrest in genes downstream the Wnt and Notch pathways. This expression pattern precedes the events leading to the maturation stage within PSM cells. Consequently, our model implies that an explicit maturation wavefront does not need to be assumed, but instead, it is an emergent property of the somitogenesis regulatory network contained within each PSM cell (see Fig. 5).

- The transition from the oscillatory regime to the high-expression state cannot be reverted by increasing the bifurcation parameter $k$ back from the critical value (0.44), as long as it remains below a higher threshold (0.76). This behavior ensures an irreversible commitment of PSM cells to somitic cells, in agreement with the experimental evicence.

- Finally, our model reproduces the opposed graded distribution of Fgf8 and RA along the PSM. The coexistence of an oscillatory behavior in all three signaling pathways and of the opposed Fgf8/RA gradients is, to our knowledge, a characteristic that has not been previously described by other model.

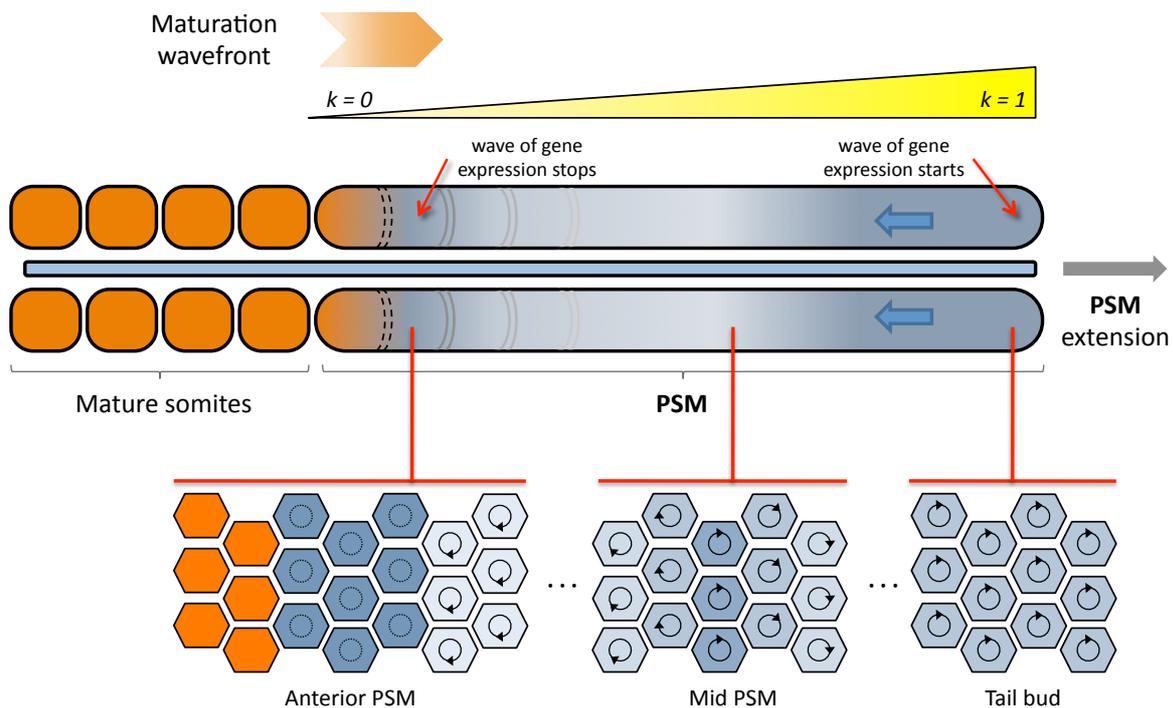

**Fig. 5** Spatiotemporal control of the segmentation process. The bifurcation parameter $k$ controls the oscillation period, amplitude and final arrest along the PSM. A somitic (mature) state is shown in orange, whereas a presomitic (immature) state is shown in blue. The dark blue at the anteriormost PSM denotes arrested oscillations, implying that those cells have become committed to somitic and are able to start their maturation stage. Differences in the oscillation phase of adjacent cells located at mid PSM and tail bud are shown with different blue intensities and inner edged circles as predicted by the simulation in Movie S1 (see Supporting Material).

In summary, we have introduced a simplistic model, which nonetheless is consistent with the global regulatory network underlying amniotic somitogenesis, and have proved that it can reproduce, with a proper parameter choice, most of the experimentally-observed dynamical features of this phenomenon. Of particular interest is the suggestion that oscillation arrest takes place via an infinite period bifurcation, arising from the interaction of a limit cycle (originated by a delayed negative feedback loop) and a bistable switch (originated by a positive feedback circuit). As a matter of fact, our model still reproduces an infinite period bifurcation when the FGF, Fgf8, and RA nodes are removed from the network, but the hysteretic behavior is lost. In other words, although we have given a higher weight to the Notch/Wnt interactions and they are sufficient to explain the arrest of oscillations, the rest of the network elements are necessary to explain other important behaviors, like hysteresis.

The model regulatory-vector signs (Fig. 1) are supported by biological evidence (see Supporting Table S1). However, since the detailed wiring of the somitogenesis regulatory network is much more complex than that portrayed in Fig. 1, other architectures are possible. In fact, we explored some of them but were not able to find another one capable of resembling the system dynamics as well as the one here reported does. We also tested parameter sets that emphasized different interactions of the network sketched in Fig. 1, but were only able to reproduce the reported behavior when the parameter values are chosen as described in the Results section. Of course, this empirical search is far from being conclusive, and it would be interesting to verify whether different network architectures and different parameter ranges are compatible with the system dynamic behavior. Nevertheless, this falls beyond the scope of the present paper.